# Visible light emission from a silica microbottle resonator by second and third harmonic generation

M. Asano,[1] S. Komori,[1] R. Ikuta,[1] N. Imoto,[1] Ş. K. Özdemir,[2,*] and T. Yamamoto[1,*]

[1] Department of Materials Engineering Science, Graduate School of Engineering Science, Osaka University, Toyonaka, Osaka 560-8531, Japan
[2] Department of Electrical and Systems Engineering, Washington University, St. Louis, MO 63130, USA

*Corresponding author: ozdemir@wustl.edu; yamamoto@mp.es.osaka-u.ac.jp

**We report the first observation of nonlinear harmonic generation and sum frequency generation (SFG) coupled with stimulated Raman scattering (SRS) via the second-order ($\chi^{(2)}$) and the third-order ($\chi^{(3)}$) nonlinearities in a silica microbottle resonator. The visible light emission due to third harmonic generation (THG) was observed in both the output of a tapered fiber and the optical microscope images, which can be used to identify the axial mode profiles. SFG enabled by three- and four-wave mixing processes between the pump light and the light generated via SRS was also observed. Second Harmonic generation (SHG) and the SFG are enabled by $\chi^{(2)}$ induced in silica by surface effects and multipole excitations.**

http://dx.doi.org/10.1364/OL.99.099999

Nonlinear harmonic generation and sum frequency generation (SFG) in micro- and nano- photonic devices have attracted increasing interest due to the possibility of developing light sources for spectral bands for which currently there is no available laser or laser material, and for their use as wavelength converters required for distributed quantum and classical communication and information processing [1]. Fibers, waveguides and resonators fabricated from silica which has third order optical nonlinearity $\chi^{(3)}$ and very low loss in the telecommunication bands, are widely used to obtain visible light from pump laser(s) in the telecommunication band through third harmonic generation (THG) or SFG enabled by four-wave mixing (FWM) [2-7]. Macroscopic inversion symmetry which forbids second order nonlinear processes in amorphous silica can be broken via poling, interface effects, surface or multipole excitations giving rise to a permanent second order optical nonlinearity $\chi^{(2)}$ [8-10]. As a result, processes such as second harmonic generation (SHG), three-wave mixing (TWM) and sum frequency generation (SFG) become possible in silica fibers [11-13]. Another widely used approach for SHG in silica microresonators is to coat the resonator with a nonlinear optical material with $\chi^{(2)}$ [14].

Optical resonators have been used to enhance nonlinear interactions because they can confine light in small mode volumes (V) for long times (high quality factors Q), leading to increased field intensity. As a result power budget required for the nonlinear process is strongly lowered. Third harmonic generation has been demonstrated in droplets [15-17], silica microspheres [7], silica microtoroids [6], photonic crystal waveguides [18], and silicon nitride ($Si_3N_4$) microrings [19]. Second harmonic generation has been reported in $Si_3N_4$ microrings [19] and in silica microfibers and loop resonators [13] via $\chi^{(2)}$ induced by symmetry breaking due to asymmetric dipole potential formed at the interfaces (interface between $Si_3N_4$ core and $SiO_2$ clad in $Si_3N_4$ microrings, and glass-air interface in silica microfibers and loops) and bulk multipole contributions.

Whispering-gallery-mode microresonators fabricated in various shapes with many different materials have served as a very promising platform for nonlinear optics. In addition to the conventional nonlinear processes such as SHG [19-23], THG [6,7,22], SFG [6,7,24,25], high-Q WGM microresonators have been shown to support coupled nonlinear interactions, such as SFG-coupled stimulated Raman scattering (SRS) [6,7] and the FWM-coupled stimulated Brillouin scattering (SBS) [26], when they are designed with proper spectral mode distribution (i.e., doubly or triply resonant, or with free-spectral-range (FSR) precisely matching the Brillouin frequency shift, etc.) and are driven with sufficiently high pump power.

Recently solid-core microbottle resonators fabricated from single mode silica optical fibers (125 μm clad and 10 μm core) using a heat-and-pull method have attracted much interest because they are easy to fabricate, have high optical quality factors (Q~$10^8$), and confine light in 3D [27,28]. A unique property of microbottle resonators is that they have WGMs both in the azimuthal and axial directions. Thus, their FSR can be engineered via the axial length of the bottle, which enables to fabricate resonators with the desired density of resonances so that the resonances are located at the frequencies of interest, without changing the micro-scale radius of the bottle. This makes microbottles attractive for studying nonlinear processes, in particular for those requiring doubly or triply resonant cavities.

Up to date, solid-core silica microbottles have been used in optomechanics [29], cavity-QED [30], add-drop filter [31], Kerr switch [32], and slow-light device [33]. In this letter, we report the first observations of second and third harmonic generation (SHG and THG) and stimulated Raman scattering coupled sum frequency generation (SRS-SFG) in a silica microbottle resonator.

The nonlinear processes that we have observed for the first time in a silica solid-core microbottle resonator are depicted in Fig. 1(a). The third harmonic is generated due to $\chi^{(3)}$ of silica via FWM whereas the second harmonic is generated via TWM enabled by $\chi^{(2)}$ induced due to interface and bulk multipoles. The SRS-SFG originates from the nonlinear interaction enabled by $\chi^{(3)}$ or $\chi^{(2)}$ between the pump light and the Stokes light generated via SRS. An image of the silica microbottle resonator used in our experiment is shown in Fig. 1(b). The diameter of the microbottle reaches to 64.5 μm at its largest region, and the two necks of the bottle, which are 3.5 mm away from each other, have the diameter of 23.1 μm. A schematic illustration of the experimental setup is depicted in Fig. 1(c). Light from an external cavity diode laser (ECDL) with a wavelength in the 1550 nm band and linewidth less than 300 kHz was amplified by an erbium-doped fiber amplifier (EDFA) and then coupled into WGM of the microbottle resonator via tapered fiber to excite nonlinear processes. The frequency of the pump light was scanned linearly around a resonance using a triangle signal generated from an arbitrary function generator (AFG). The transmittance of the tapered fiber was 95.3 % for the pump laser in the 1550 nm band. The transmission spectra of the microbottle were monitored by a photodetector (PD) connected to a digital sampling oscilloscope (DSO) to probe the WGM resonances and measure their quality factors. The light generated via the nonlinear processes was measured using optical spectrum analyzers OSA1 (for wavelength longer than 600 nm) and OSA2 (for wavelengths shorter than 900 nm). An optical microscope connected to a PC was also used to directly monitor the visible light emission from the microbottle.

The THG process occurs among the resonant modes satisfying the phase matching condition (i.e. energy conservation and momentum conservation) expressed as $\omega_{\text{THG}} = 3\omega_{\text{Pump}}$ and $\mathbf{k}_{\text{THG}} = 3\mathbf{k}_{\text{Pump}}$ where $\omega_i$ and $\mathbf{k}_i$ are the angular frequency and the wavevector, respectively. Figure 2 shows the results of the characterization of THG signal in the silica microbottle resonator. In our experiments, the tapered silica fiber and the taper-resonator coupling were optimized for the pump mode which was in the 1550 nm band (i.e., IR light). Thus, the third-harmonic signal (i.e., green light) detected at the output of the fiber was very weak (i.e., barely above the noise level at low pump powers), because under

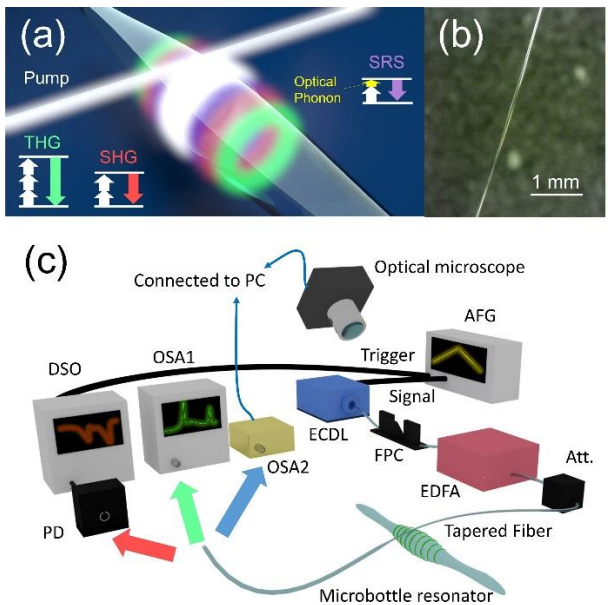

**FIG. 1** (a) An illustration of the nonlinear processes observed in silica microbottle resonator. THG: Third harmonic generation; SHG: Second harmonic generation, SRS: Stimulated Raman scattering. (b) Optical microscope image of the microbottle resonator used in the experiment. (c) Schematic illustration of the experimental setup. ECDL: External cavity diode laser; FPC: Fiber-based polarization controller; EDFA: Erbium-doped fiber amplifier; Att.: Attenuator; AFG: Arbitrary function generator; PD: photodetector; DSO: Digital sampling oscilloscope; OSA1 & OSA2: Optical spectrum analyzer.

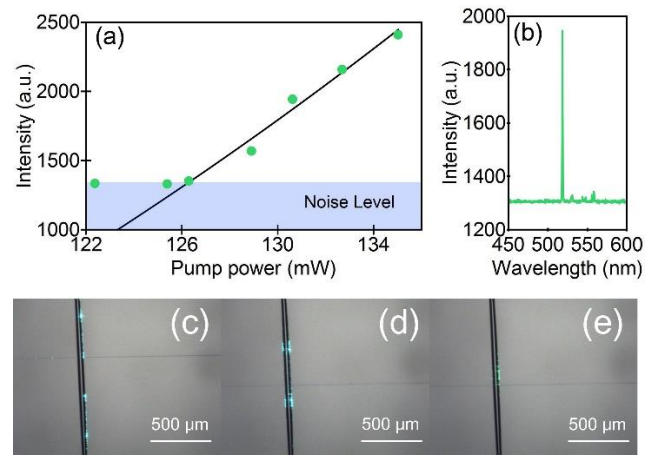

**FIG. 2** Third harmonic generation (THG) in a silica microbottle resonator. (a) Pump power versus intensity of the third harmonic signal. Blue shaded area corresponds to the detector noise level in which spectra could not be observed with the OSA2. (b) Third harmonic spectrum at 519.1 nm obtained with the pump light of wavelength 1557.2 nm at a pump power of 130.6 mW. (c), (d) and (e) Axial mode profiles of THG signal in the microbottle resonator with pump lights of power (c) 123.2 mW, (d) 106.3 mW, and (e) 102.9 mW, at wavelengths (c) 1562.1 nm (d) 1558.7 nm and (e) 1565.2 nm. Emission in (e) is due to the stimulated Raman scattering coupled sum frequency generation (SRS-SFG) and hence it has a different color than the emissions in (c) and (d). Note that the emission patterns in Figs. 2(c)-(e) are obtained at different positions of the tapered fiber along the axial direction of the microbottle resonator while keeping the system at critical coupling.

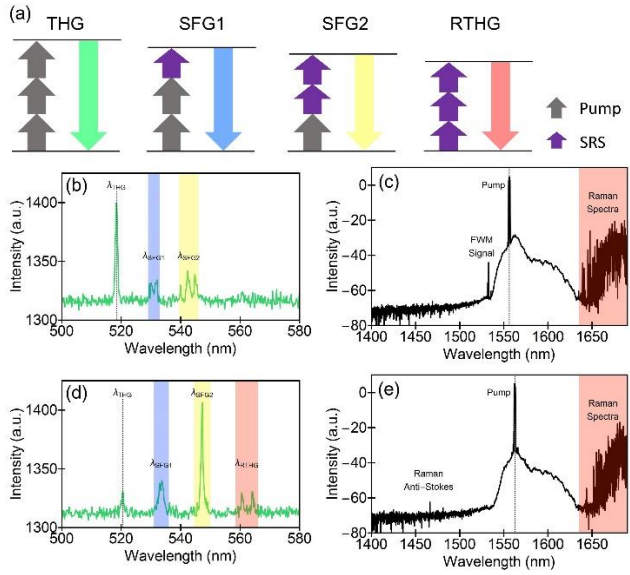

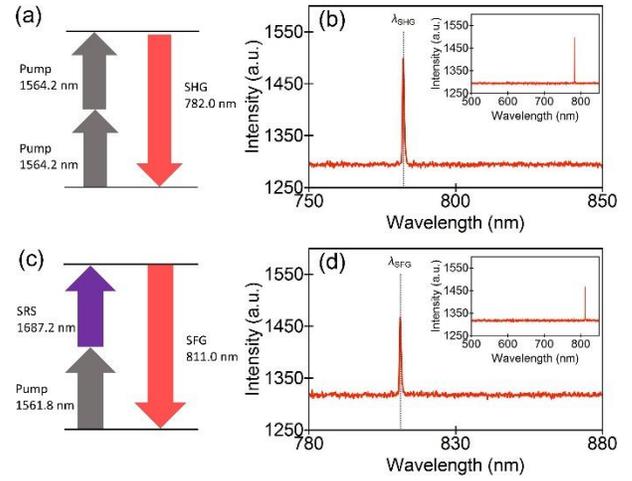

**Fig. 3** Nonlinear processes observed in silica microbottle resonator due to $\chi^{(3)}$. (a) Energy diagram of THG generation from pump, SFG by mixing of two pump and one Raman photon (SFG1) and of one pump and two Raman photons (SFG2), and THG from Raman light generated by SRS (RTHG). The optical spectra of visible light generated by these processes are shown in (b) and (d). The optical spectra obtained for pump in the IR band show the generation of light via SRS in (c) and (e), generation of signal and idler (parametric oscillation) via FWM in (c) and the FWM-coupled SRS in (e). The wavelength and the power of the pump light was 1556.1 nm and 208.3 mW in (a) and (b), and 1562.3 nm and 171.0 mW in (c) and (d).

this condition the shorter evanescent tail of green light does not have optimal overlap with the taper for efficient extraction of green light from the resonator. Moreover, the silica taper fiber has larger absorption in the visible band, which further decreasing the amount of green light reaching to the detectors. Moving the tapered fiber closer to the resonator and hence bringing the taper-resonator coupling into deep-overcoupling regime allowed a larger overlap between the evanescent tail of the green light and the mode of the tapered fiber. However, this shifted the pump mode from its optimal condition, which we compensated by using a higher pump power. With a pump laser at wavelength 1557.2 nm and with a power of 126.3 mW, the third harmonic signal observed at the wavelength of 519.1 nm increased above the noise level, and further increased with increasing pump power (see Fig.2 (a)).

By bringing the taper-resonator coupling to critical coupling for the pump mode (i.e., 1550 nm band), we monitored the generation of third harmonic signal directly with an optical microscope placed over the microbottle. In this coupling condition, intracavity field intensity for the pump mode reaches its maximum with efficient pumping, and the escape of the third harmonic signal into the taper is minimal, allowing the observation of bright third harmonic signal as seen in Fig. 2(c) and (d). This allowed to image axial mode profile of the microbottle resonator as seen in the optical microscope images obtained at different pump powers and wavelengths (Figs. 2(c)-(e)). Two different axial modes covering over several millimeters were clearly observed (Fig. 2 (c) and (d)). The emission shown in Fig. 2(e) is the result of SFG induced by the nonlinear mixing of the pump and the light generated via SRS. This emission has a different color and its axial mode spanned a shorter distance with less spacing than the ones seen in Figs. 2(c) and (d).

**Fig. 4** Second harmonic (SHG) and sum frequency generation (SFG) via three-wave mixing due to $\chi^{(2)}$ in a silica microbottle. (a) and (c) Energy diagrams of the SHG and the SFG process, respectively. (b) The emission via SHG had a wavelength of 782.0 nm for a pump at 1564.2 nm. (d) The emission via SFG had a wavelength at the 811.0 nm for a pump with wavelength 1561.8 nm. The insets of (b) and (d) show the whole spectrum in the visible band, implying that nonlinear processes due to third-order nonlinearity did not take place in these experiments.

Energy diagrams of possible routes to generate different colors of light via FWM process in a silica microbottle resonator is depicted in Fig. 3(a). In addition to THG from the pump light and THG from the Raman light generated via SRS that we named as Raman third harmonic generation RTHG, there are two SFG pathways. In SFG1, two pump photons and one SRS photon mix to create a photon with a higher energy if the phase matching condition described as $\omega_{\text{SFG1}} = 2\omega_{\text{Pump}} + \omega_{\text{SRS}}$ (energy conservation) and $\boldsymbol{k}_{\text{SFG1}} = 2\boldsymbol{k}_{\text{Pump}} + \boldsymbol{k}_{\text{SRS}}$ (momentum conservation) is satisfied. In SFG2, on the other hand, one pump photon and two SRS photons mix to create a photon with a higher energy if the phase matching condition $\omega_{\text{SFG2}} = \omega_{\text{Pump}} + 2\omega_{\text{SRS}}$ and $\boldsymbol{k}_{\text{SFG2}} = \boldsymbol{k}_{\text{Pump}} + 2\boldsymbol{k}_{\text{SRS}}$ is fulfilled.

By scanning the laser frequency and changing the center wavelength and the power of pump light, we obtained a variety of nonlinear optical spectra. In addition to the standard THG form the pump light, we observed two types of SFG spectra in the visible band with a pump of wavelength 1556.1 nm and of power of 208.3 mW (Fig. 3 (b)). Scanning the pump frequency and using a stronger pump power induced SRS combs that are separated about 100 nm from the pump spectrum (Fig. 3(c)). In the SFG process, the single or multiple SRS spectra in this comb were extracted such that the resonant pump modes and the SRS spectra satisfied the phase matching condition. As a result, we observed emissions at different wavelengths in the visible band as shown in shaded regions of Fig. 3(b). We also observed parametric oscillation due to FWM with strong pump power as evidenced by the presence of a signal peak in the pump band (Fig. 3(c)). We could not resolve the idler in the spectra. One possible reason for this may be that the process is SRS assisted non-degenerate FWM [34] in which case the idler is located in the region labeled as Raman spectra. Another reason may be the amplified spontaneous emission from EDFA which introduces large noise.

With a pump of wavelength 1562.3 nm and power of 171.0 mW, we observed all the processes shown in Fig. 3(a): THG from pump, THG from the Raman emission (RTHG), SFG1 and SFG2. The spectra of the light generated by these processes are given in Fig.3 (d). The emission peak labeled as Raman anti-Stokes in the IR band in Fig. 3(e) was due to FWM coupled SRS process which takes place when the phase matching condition $\omega_{\text{Anti-SRS}} = 2\omega_{\text{Pump}} - \omega_{\text{SRS}}$ and $\boldsymbol{k}_{\text{Anti-SRS}} = 2\boldsymbol{k}_{\text{Pump}} - \boldsymbol{k}_{\text{SRS}}$ is satisfied.

Finally, we report the observation of SHG in our silica microbottle resonator. SHG and SFG via three-wave mixing (TWM) should satisfy the energy conservation $\omega_{\text{SHG}} = 2\omega_{\text{Pump}}$ and $\omega_{\text{SFG}} = \omega_{\text{Pump}} + \omega_{\text{SRS}}$, and the momentum conservation $\boldsymbol{k}_{\text{SHG}} = 2\boldsymbol{k}_{\text{Pump}}$ and $\boldsymbol{k}_{\text{SFG}} = \boldsymbol{k}_{\text{Pump}} + \boldsymbol{k}_{\text{SRS}}$ (Fig.4 (a) and (c)). We observed light emission from silica resonators via SHG and SFG when the taper-resonator coupling was set to deep overcoupling regime to fix the coupling condition for the CW pump laser (in 1550 nm band) whose frequency was finely tuned and thermally locked to the resonance mode. Figure 4 (b) and (d) show the SHG signal and the SFG signal generated via TWM between the pump light and the light generated via SRS. The SHG signal at the wavelength of 782.0 nm and the SFG signal at the wavelength of the 811.0 nm were obtained at a pump power of 209.4 mW when the pump wavelength was 1564.2 nm and 1561.8 nm, respectively. We confirmed that under these conditions, THG or SFG via FWM did not occur in our system, as seen in the extended spectra shown in the insets of Figs. 4(b) and (d). We note that our experiments were performed with CW pump lasers. Compared to the previous reports on SHG in silica fibers and fiber loop resonator [13] with strong optical pulses, we find that the pump power in our CW laser was 430 times smaller than the peak power of the pulses used for SHG in fiber loop resonators.

In summary, we report the first observation of nonlinear harmonic generation and SRS-coupled SFG due to $\chi^{(3)}$ and $\chi^{(2)}$ in a silica microbottle resonator. The THG in the microbottle resonator distributed over a millimeter-scale along the axial direction is a unique property seen only in these resonators. Finally we believe that interface and surface effects play a significant role in the emergence of $\chi^{(2)}$ in our silica microbottle resonators because we have not used poling or dopants during the fabrication of our resonators. Extension of our observations to other types of WGM silica microresonators such as microdisk, microsphere or microtoroids may require poling or engineered interface and optimized dimension. Our demonstration of SFG both for FWM and TWM can be also extended to quantum wavelength conversion [35-37] by decreasing the pump power to remove the noises and by injecting single photons instead of the SRS signal.

**Funding.** This work was supported by MEXT/JSPS KAKENHI Grant Number JP16H01054, JP16H02214, JP15H03704 and JP15KK0164. S.K.O is supported by ARO Grant No. W911NF-16-1-0339.

**Acknowledgment**. M.A acknowledges the support from Program for Leading Graduate Schools: "Interactive Materials Science Cadet Program".